\documentclass[11pt]{iopart}

\usepackage{iopams}
\usepackage{graphicx}
\usepackage{setspace}
\usepackage[dvips,usenames]{color}
\usepackage{fancyhdr}
\scrollmode
\eqnobysec

\begin{document}

\fancyhf{}
\fancyhead[L]{\textit{\nouppercase{Distribution of the least-squares estimators}}}
\fancyhead[R]{\nouppercase{D. Boyer} \textit{\nouppercase{et al}}}
\fancyfoot[C]{\thepage}

\title[Distribution of the least-squares estimators]{Distribution of the least-squares estimators of a single Brownian trajectory diffusion coefficient}

\author{Denis Boyer$^1$, David S Dean$^2$,
Carlos Mej\'{\i}a-Monasterio$^{3,4}$ and Gleb Oshanin$^5$}

\address{$^1$ Instituto de F\'{\i}sica, Universidad Nacional Autonoma de Mexico,
D.F. 04510, Mexico}
\address{$^2$ Universit\'e de  Bordeaux and CNRS, Laboratoire Ondes et
Mati\`ere d'Aquitaine (LOMA), UMR 5798, F-33400 Talence, France}
\address{$^3$ Laboratory of Physical Properties,
Technical University of Madrid, Av. Complutense s/n, 28040 Madrid, Spain}
\address{$^4$Department  of Mathematics and  Statistics, University  of 
Helsinki,  P.O.   Box  68  FIN-00014  Helsinki,  Finland}
\address{$^5$Laboratoire de Physique Th\'eorique de la Mati\`ere Condens\'ee (UMR CNRS 7600),
Universit\'e Pierre et Marie Curie/CNRS, 4
place Jussieu, 75252 Paris Cedex 5 France}

\eads{\mailto{boyer@fisica.unam.mx}, \mailto{david.dean@u-bordeaux1.fr}, \mailto{carlos.mejia@upm.es}, \mailto{oshanin@lptmc.jussieu.fr}}

\begin{abstract}
  In this paper we  study the distribution function $P(u_{\alpha})$ of
  the estimators $u_{\alpha} \sim T^{-1} \int^T_0 \, \omega(t) \, {\bf
    B}^2_{t} \,  dt$, which optimise the least-squares  fitting of the
  diffusion  coefficient $D_f$  of a  single  $d$-dimensional Brownian
  trajectory ${\bf  B}_{t}$.  We pursue here  the optimisation further
  by considering a family of weight functions of the form $\omega(t) =
  (t_0 + t)^{-\alpha}$,  where $t_0$ is a time lag  and $\alpha$ is an
  arbitrary real number, and seeking such values of $\alpha$ for which
  the  estimators most  efficiently filter  out the  fluctuations.  We
  calculate  $P(u_{\alpha})$   exactly  for  arbitrary   $\alpha$  and
  arbitrary spatial dimension $d$, and show that only for $\alpha = 2$
  the distribution $P(u_{\alpha})$ converges, as $\epsilon = t_0/T \to
  0$,  to the Dirac  delta-function centered  at the  ensemble average
  value of  the estimator.  This allows  us to conclude  that only the
  estimators with  $\alpha = 2$  possess an ergodic property,  so that
  the ensemble averaged diffusion coefficient can be obtained with any
  necessary precision from a single trajectory data, but at the expense of
  a  progressively higher  experimental resolution.   For  any $\alpha
  \neq 2$  the distribution  attains, as $\epsilon  \to 0$,  a certain
  limiting  form with  a finite  variance, which  signifies  that such
  estimators are not ergodic.
\end{abstract}

\vspace{2pc}

\submitto{JSTAT}
\maketitle

\tableofcontents

\section{Introduction}

Single particle tracking (SPT) is an increasingly used method of analysis in biological and 
sot matter systems where the trajectories of individual particles can be optically observed. Recent advances in image processing, data storage and microscopy have led to an increasing number of 
papers, notably in biophysics, on single particle tracking in biological settings such as the cellular interior and the cell membrane. However the basic method of SPT owes its origin to the work of Perrin on Brownian motion \cite{perrin}, where optical observation is used to 
generate the  time series for the position of an individual particle trajectory ${\bf B}_t$ in a medium (see, e.g., Refs. \cite{bra,saxton}). Complemented by the appropriate theoretical analysis,
the information drawn from a single, or a finite number of trajectories,
provides insight into the underlying physical mechanisms governing the transport properties of the particles. Via the analysis of the stochastic processes manifested in single particle trajectories, SPT is routinely used for  the microscopic characterisation of the thermodynamic
rheological properties of complex media \cite{mason},
and also to identify non-equilibrium biological effects, for example the
motion of biomolecular motors \cite{greenleaf}. In biological cells
and complex fluids, SPT methods have
 become instrumental in demonstrating deviations from normal
Brownian motion of passively moving particles (see, e.g.,
Refs.\cite{golding,weber,bronstein,seisenberger,weigel}).
The method is thus potentially a
powerful tool to probe physical and biological processes at the level
of a single molecule \cite{moerner}.

The reliability  of the information drawn from  SPT analysis, obtained
at high temporal and spatial  resolution but at expense of statistical
sample size is not  always clear.  Time averaged quantities associated
with    a    given    trajectory    may   be    subject    to    large
trajectory-to-trajectory fluctuations.  For  a wide class of anomalous
diffusions, described  by continuous-time random  walks, time-averages
of  certain   particle's  observables  are,  by   their  very  nature,
themselves  random  variables distinct  from  their ensemble  averages
\cite{rebenshtok,ralf}.   For   example,   the   square   displacement
time-averaged  along  a given  trajectory  differs  from the  ensemble
averaged   mean   squared   displacement\cite{ralf,he,lubelski}.    By
analyzing  time-averaged  displacements  of  a  particular  trajectory
realization,  subdiffusive motion can  actually look  normal, although
with strongly differing diffusion  coefficients from one trajectory to
another  \cite{ralf,he,lubelski} and  show pronounced  ageing effects
\cite{barkai}.

Standard  Brownian  motion  is  a  much simpler  random  process  than
anomalous diffusion,  however the analysis of its  trajectories is far
from being  as straightforward as one  might think, and  all the above
mentioned troublesome problems exist for Brownian motion as well. Even
in     bounded      systems,     substantial     manifestations     of
trajectory-to-trajectory fluctuations in  first passage time phenomena
have been revealed \cite{carlos,thiago}.
If one is interested in determining the diffusion coefficient of a
given  particle,
standard  fitting  procedures  applied  to  finite  albeit  very  long
trajectories  unavoidably  lead to  fluctuating
estimates  $D_f$ of  the diffusion  coefficient, which  might  be very
different  from the  true ensemble  average  value $D$,  defined in  a
standard fashion as \vspace{1pc} \begin{equation}
\label{msd}
D = \frac{\mathbb{E}\left\{{\bf B}^2_t\right\}}{2 d t} \,,
\end{equation}\vspace{1pc}  
where the symbol $\mathbb{E}\{\ldots\}$ denotes the ensemble average and
$d$ is the spatial dimension. 

To get a rough idea of how basic estimators for diffusion constants can fluctuate, 
consider a simple-minded, rough  estimate of $D_f$,
defining it as  the slope of the line connecting  the starting and the
end-points  ${\bf B}_t$  of  a  given trajectory,  i.e.,  like $D$  in
Eq.~(\ref{msd}) but without averaging, that  is $D_f = {\bf B}^2_t/2 d
t$. A  single trajectory diffusion  coefficient $D_f$ so defined  is a
random variable  whose probability density function  (pdf) $P(D_f)$ is
the so-called  chi-squared distribution  with $d$ degrees  of freedom:
\vspace{1pc} \begin{equation}
\label{chi-s}
P(D_f) = \frac{1}{\Gamma(d/2)} \, \left(\frac{d}{2 D}\right)^{d/2} \, D_f^{d/2 - 1} \, \exp\left(- \frac{d}{2} \cdot \frac{D_f}{D} \right) \,,
\end{equation}\vspace{1pc}  
where  $\Gamma(\cdot)$ is the  Gamma-function. The  pdf in  the latter
equation diverges as $D_f \to 0$ for $d=1$, $P(D_f=0)$ is constant for
$d=2$, and  only for $d >  2$ the distribution has  a bell-shaped form
with the most  probable value $D_f^* = (1 - 2/d)  D$. This means that,
e.g., for $d=3$, it is most likely that extracting $D_f$ from a single
Brownian trajectory  using this simple-minded approach  we will obtain
just one third of the true diffusion coefficient $D$.

  As a  matter of  fact, even taking  advantage of  more sophisticated
  fitting  procedures, variations  by  orders of  magnitude have  been
  observed  in  SPT  measurements  of the  diffusion  coefficient  for
  diffusion  of  the  LacI   repressor  protein  along  elongated  DNA
  \cite{austin}, in the plasma membrane \cite{saxton} or for diffusion
  of a  single protein in  the cytoplasm and nucleoplasm  of mammalian
  cells \cite{goulian}.  Such a broad  dispersion of the values of the
  diffusion  coefficient   extracted  from  SPT   measurements  raises
  important questions about the optimal methodology, much more robust than
  the rough estimate  mentioned above, that should be used
  to  determine the true  ensemble average  value of  $D$ from  just a
  single trajectory.
Clearly, it is highly desirable  to have a reliable estimator even for
the  hypothetical pure  cases, such  as, e.g.,  unconstrained standard
Brownian motion. A reliable estimator must possess an ergodic property
so  that its  most  probable  value should  converge  to the  ensemble
average one  and the  variance should vanish  as the  observation time
increases. This  is often not the  case and moreover,  ergodicity of a
given estimator  is not known \textit{a  priori} and has  to be tested
for  each  particular form  of  the  estimator.   On the  other  hand,
knowledge of  the distribution  of such an  estimator could  provide a
useful gauge to  identify effects of the medium  complexity as opposed
to  variations in  the  underlying thermal  noise driving  microscopic
diffusion. Recently, much effort has  been invested in the analysis of
this  challenging  problem and  several  important  results have  been
obtained  for the  estimators based  on the  time-averaged mean-square
displacement   \cite{greb1,greb2,greb3},    mean   maximal   excursion
\cite{tej}   or   the  maximum   likelihood   approximation  and   its
ramifications \cite{berglund,m1,m2,boyer,boyer1}.

Let us define the dimensionless estimator of the diffusion
coefficient as $u\equiv D_f/D$.
In this  paper, following our  succinct presentation in  \cite{we}, we
focus on a family of 
least-squares\footnote{This term will be made
  clear in Section \ref{LS}} estimators $u_{\alpha}$ given by
\vspace{1pc}
\begin{equation}
\label{u}
u_{\alpha} = \frac{A_{\alpha}}{T} \int^{T}_{0} \, \omega(t) \, {\bf B}^2_t \, dt,
\end{equation}\vspace{1pc}  
where $\omega(t)$ is the weight function of the form
\vspace{1pc}  \begin{equation}
\label{omega}
\omega(t) = \frac{1}{(t_0 + t)^{\alpha}} \,,
\end{equation}\vspace{1pc}  
$\alpha$ being a  tunable exponent, (positive or negative),  $t_0$ - a
lag time and $A_{\alpha}$  - the normalisation constant, appropriately
chosen such that $\mathbb{E}\left\{u_{\alpha}\right\} \equiv
1$. Therefore \vspace{1pc} \begin{equation}
\label{aalpha}
A_{\alpha} = \frac{T}{2 d D} \left(\int^T_0 \frac{t \, dt}{(t_0 + t)^{\alpha}}\right)^{-1}\,.
\end{equation}\vspace{1pc}  
Note  that such  a normalisation  permits a  direct comparison  of the
effectiveness  of  estimators  corresponding  to different  values  of
$\alpha$.   Our  goal  here  is   to  find  such  $\alpha$  for  which
$u_{\alpha}$  is  ergodic, namely, so  that  the  single trajectory  
diffusion coefficient $D_f \to D$ (or $u_{\alpha}\rightarrow 1$) 
as $\epsilon = t_0/T \to 0$.

It should be emphasised that, as a matter of fact, we are dealing here
with two  consecutive optimisation  schemes: first, the  estimators in
Eq.~(\ref{u})  are  already the  results  of  an  optimisation of  the
least-squares fitting procedure for the diffusion coefficient $D_f$ of
a single Brownian trajectory  and second, an optimisation is performed
for  the  weight  function  $\omega(t)$  chosen  from  the  family  of
functions in Eq.~(\ref{omega}).

This paper is outlined as follows: We start in Section \ref{LS} with a
physical interpretation  of the  estimators given in Eq.~(\ref{u})  and show
that  these  stem  out   of  the  minimisation  procedure  of  certain
least-squares  functionals of  the square displacement ${\bf B}^2_t$.  
Next, in  Section
\ref{basic} we  introduce basic notations  and the definitions  of the
characteristic properties we are  going to study. Section \ref{mgf} is
devoted  to the evaluation  of the  moment-generating function  of the
estimators. In Section \ref{variance}  we present explicit results for
the variance of  the least-squares estimators for $\alpha  \neq 2$ for an
infinite observation time, the  variance for the case $\alpha = 2$
for arbitrary observation time. We also discuss  the optimisation  of the
variance   of   the   least-squares   estimators  in   the   case   of
continuous-time and -space  Brownian trajectories recorded at discrete
time moments.   Next, in Section \ref{asymp} we  obtain the asymptotic
behaviour  of the  distribution $P(u_{\alpha})$  in  arbitrary spatial
dimension. Further  on, Section \ref{dist} discusses the shape of the
full distribution $P(u_{\alpha})$ and the location of  the most probable 
values of  the estimators for three and  two-dimensional  systems. We also
study the distribution of the variable
$\omega_{\alpha}\equiv u^{(1)}_{\alpha}/(u^{(1)}_{\alpha}+u^{(2)}_{\alpha})$, 
with $u^{(1)}_{\alpha}$ and $u^{(2)}_{\alpha}$ two independent estimates.
Finally,  in  Section  \ref{conc}  we
conclude with a brief summary of our results.

\section{Physical interpretation of the estimators $u_{\alpha}$.}
\label{LS}

Before we proceed, it might  be instructive to understand where do the
functionals   in   Eq.~(\ref{u})   stem   from   and   what   physical
interpretation  may  they  have.   Consider  a  given  $d$-dimensional
trajectory ${\bf  B}_t$ with  $t \in [0,T]$  and try to  calculate the
diffusion  coefficient  $D_f$  of  this  given  trajectory  using  the
least-squares approximation for the whole trajectory. To this purpose,
one     writes      the     following     least-squares     functional
\vspace{1pc} \begin{equation}
\label{func}
F = \frac{1}{2} \int^T_0 \frac{\omega(t)}{t} \, \left({\bf B}^2_t - 2 d D_f t\right)^2 \,dt \, ,
\end{equation}\vspace{1pc}  
and  seeks  to  minimise  it  with  respect to  the  value  of  $D_f$,
considered as a variational  parameter.  Note that Eq.~(\ref{func}) is
a  bit  more  general  compared  to  the  usually  used  least-squares
functionals.  A novel feature here is that in Eq.~(\ref{func}) we have
introduced a  {\it weight} function $\omega(t)$  which, depending on
whether  it is a  decreasing or  an increasing  function of  $t$, will
sensitive to the short  time or the long time  behavior of the trajectory
${\bf B}_t$, respectively.

Furthermore, 
setting the  functional derivative $\partial F/\partial  D_f$ equal to
zero,   we  find   that   the   minimum  of   $F$   is  obtained   for
\vspace{1pc}   \begin{equation}   \frac{D_f}{D}  =   \left(\frac{1}{T}
    \int^T_0  dt \,  \omega(t) \,  {\bf  B}^2_t\right)/\left(\frac{2 d
      D}{T} \int^T_0 dt \, t \, \omega(t) \right) \,.
\end{equation}\vspace{1pc}  
Next, identifying  the denominator with $1/A_{\alpha}$ in Eq.~(\ref{aalpha}), we
conclude
that $u_{\alpha}$ in Eq.~(\ref{u}) \textit{minimises} the least-squares functionals
with a weight function $\omega(t) = (t_0 + t)^{-\alpha}$.

It  is  interesting  to  note  that with  this  weight  function,  the
functional (\ref{u})  interpolates two  well known estimators  for the
diffusion  constant: In  the case  of $\alpha  = -  1$,  the estimator
$u_\alpha$  corresponds to a  usual unweighted  least-squares estimate
(LSE)      of     the      time-averaged      squared     displacement
\cite{saxton,goulian,saxton2}.  The case $\alpha = 1$ arises
in a conceptually different fitting procedure in which the conditional
probability  of   observing  the  whole  trajectory   ${\bf  B}_t$  is
maximised, subject to the constraint  that it is drawn from a Brownian
process with the mean-square  displacement $2 d D t$, Eq.~(\ref{msd}).
This is  the so-called maximum  likelihood estimate (MLE)  which takes
the  value of  $D_f$ that  maximises  the likelihood  of ${\bf  B}_t$,
defined as: \vspace{1pc} \begin{equation}  L = \prod_{t = 0}^T \left(4
    \pi  D_f  t\right)^{-d/2} \exp\left(  -  \frac{{\bf B}^2_t}{4  D_f
      t}\right) \,,
\end{equation}\vspace{1pc}  
where     the    trajectory     ${\bf    B}_t$     is    appropriately
discretized.  Differentiating the  logarithm  of $L$  with respect  to
$D_f$  and  setting  $d \ln  L/d  D_f  =  0$,  one finds  the  maximum
likelihood estimate  (see, e.g., Refs.\cite{berglund,boyer,boyer1}) of
$D_f$, which  upon a proper normalisation is  defined by Eq.~(\ref{u})
with $\alpha = 1$.

\section{Basic notations and definitions}
\label{basic}

The  fundamental  characteristic property  we  will  focus  on is  the
moment-generating  function $\Phi(\sigma)$ of  the random  variable in
Eq.~(\ref{u}): \vspace{1pc} \begin{equation}
\label{laplace}
\Phi(\sigma) = \mathbb{E}\left\{\exp\left( - \sigma u_{\alpha}\right)\right\} \,,
\end{equation}\vspace{1pc}  
where $\sigma$ is a parameter.

It  is important  to  realise  that for  standard  Brownian motion  the generating function of 
the original  $d$-dimensional problem  decomposes  into a  product of  the generating function of the 
components,  since the  squared distance  from the  origin of  a given
realisation  of   a  $d$-dimensional  Brownian  motion   at  time  $t$
decomposes into  the sum  \vspace{1pc} \begin{equation} {\bf  B}^2_t =
  \sum_{i = 1}^d B_t^2(i),
\end{equation}\vspace{1pc}  
$B_t(i)$ being realisations of trajectories of independent 1D Brownian
motions  (for  each spatial  direction).  Thus, the  moment-generating
function   $\Phi(\sigma)$  factorizes   \vspace{1pc}  \begin{equation}
  \Phi(\sigma) = G^d(\sigma),
\end{equation}\vspace{1pc}  
where
\vspace{1pc}  \begin{equation}
\label{G}
G(\sigma) = \mathbb{E}\left\{\exp\left( - \frac{\sigma A_{\alpha}}{T} \int^{T}_{0} \omega(\tau) \, B_{\tau}^2(i) \, d\tau \right)\right\}.
\end{equation}\vspace{1pc}  
In what  follows we will  thus skip the  argument $(i)$ and  denote as
$B_t$ a given trajectory of  a one-dimensional Brownian motion with an
ensemble average diffusion coefficient $D$.

The knowledge of $\Phi(\sigma)$ will  allow us to calculate directly, by
merely    differentiating   $\Phi(\sigma)$,    the    variance   ${\rm
  Var}(u_{\alpha})$ of  the distribution function  $P(u_{\alpha})$ and
to infer  the asymptotic behaviour of the  distribution function.  The
complete  distribution $P(u_{\alpha})$ will  be obtained  by inverting
the  Laplace  transform in  Eq.~(\ref{laplace})  with  respect to  the
parameter $\sigma$, namely
\vspace{1pc} 
\begin{equation}
\label{def:dist}
P(u_{\alpha}) = \frac{1}{2 \pi i} \int^{\gamma + i \infty}_{\gamma  - i \infty} d\sigma \, \exp\left(\sigma u_{\alpha}\right) \, \Phi(\sigma) \,,
\end{equation}\vspace{1pc}  
where $\gamma$ is a real number  chosen in such a way that the contour
path of integration is in the region of convergence of $\Phi(\sigma)$.
The  explicit  results  for  the  variance and  for  the  distribution
$P(u_{\alpha})$ will be presented in the next sections.

\begin{figure}[!t]
  \centerline{\includegraphics*[width=0.75\textwidth]{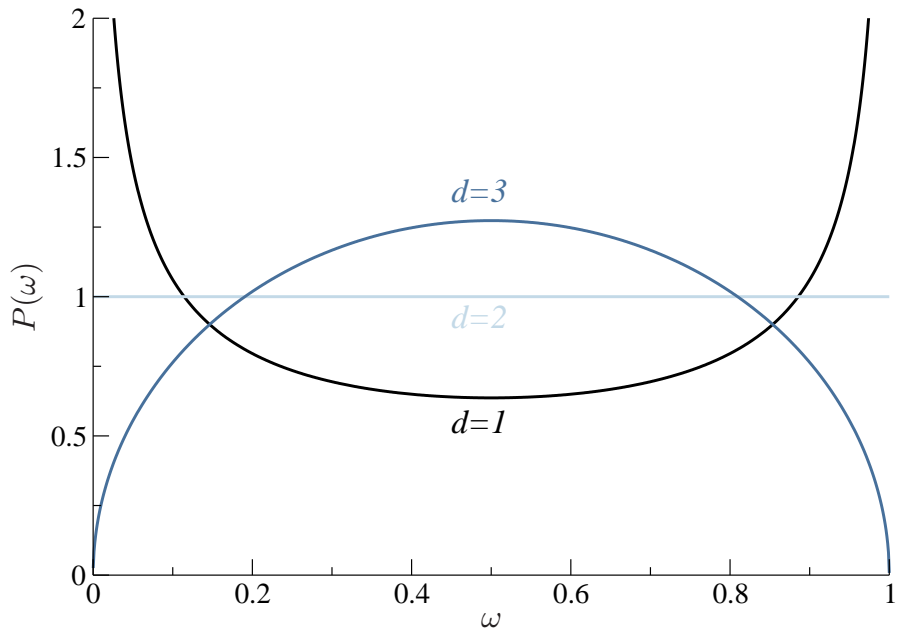}}
  \caption{(Color    online)   The    distribution    $P(\omega)$   in
    Eq.~(\ref{lim}) for $d = 1,2$ and $3$.}
\label{fig0}
\end{figure}

Further on, to highlight the role of the trajectory-to-trajectory
fluctuations, we will consider
 the probability density
  function $P(\omega_{\alpha})$ of the
random variable
\vspace{1pc}  \begin{equation}
\label{varomega}
\omega_{\alpha} = \frac{u_{\alpha}^{(1)}}{u_{\alpha}^{(1)} + u_{\alpha}^{(2)}} \,,
\end{equation}\vspace{1pc}  
where  $u_{\alpha}^{(1)}$  and  $u_{\alpha}^{(2)}$ are  two  identical
independent    random   variables    with   the    same   distribution
$P(u_{\alpha})$.   The  distribution $P(\omega_{\alpha})$,  introduced
recently       in       Ref.\cite{carlos,thiago}       (see       also
Refs.\cite{iddo,iddo1,iddo2})  is a  robust measure  of  the effective
broadness of $P(u_{\alpha})$, and \textit{probes} the likelihood
that the  diffusion coefficients  drawn from  two different
trajectories are  equal to  each other.  This  characteristic property
can   be   readily   obtained from the identity   \cite{carlos2}
\vspace{1pc} \begin{equation}
\label{romega}
 P(\omega_{\alpha}) = \frac{1}{(1 - \omega_{\alpha})^2} \, \int^{\infty}_0 du_{\alpha} \, u_{\alpha} \, P(u_{\alpha}) \,
P\left(\frac{\omega_{\alpha}}{1 - \omega_{\alpha}} u_{\alpha}\right) \,.
\end{equation}\vspace{1pc}  
Hence, $P(\omega_{\alpha})$ is known once we know $P(u_{\alpha})$. To illustrate this statement,
let us return to the simple-minded estimate of $D_f$ mentioned in the 
Introduction and the corresponding pdf given by Eq.~(\ref{chi-s}). 
In this case, $P(\omega)$ can be obtained explicitly\footnote{We drop the subscript $\alpha$ as meaningless in this case.},
\vspace{1pc}  \begin{equation}
\label{lim}
P(\omega) = \frac{\Gamma(d)}{\Gamma^2(d/2)} \, \omega^{d/2 - 1} \left(1 - \omega\right)^{d/2 - 1} \,.
\end{equation}\vspace{1pc}  
A striking feature of the beta-distribution in Eq.~(\ref{lim}) is that
its very shape depends on the spatial dimension $d$ (see
Fig.~\ref{fig0}).  For $d=1$, $P(\omega)$ is bimodal with a $U$-like
shape, the most probable values being $0$ and $1$ and, remarkably, $\omega
= 1/2$ being the least probable value.  Therefore, if we take two $1d$
Brownian trajectories, most likely we will obtain two very different
estimators of the diffusion coefficient by this method, both having 
little to do with the true ensemble average value $D$. It is
unlikely that we will get two equal values.  Further on, for $d = 2$,
$P(\omega) \equiv 1$, which signifies that \textit{any} relation
between estimates $D_f$ drawn from two different trajectories is
\textit{equally} probable.  Only for $d=3$ the pdf $P(\omega)$ is
unimodal with a maximum at $\omega = 1/2$. But even in this case it is
broad (a part of a circular arc), revealing important
trajectory-to-trajectory fluctuations. Clearly, such a simple-minded
estimate is not plausible and one has to resort to more reliable
estimators. Below we will consider the forms of $P(\omega_{\alpha})$
for the family of weighted least-squares estimators.

\section{The moment-generating function of the estimators.}
\label{mgf}

Note that the Laplace transforms of quadratic functionals of Brownian
motion (and other Gaussian processes), as the one in
Eq.~(\ref{laplace}), have attracted a great deal of interest over the
last decades following the pioneering work by Cameron and Martin
\cite{1}. Numerous results have been obtained both in the general
setting of abstract Gaussian spaces and in various specific
models generalising the original approach for Brownian motion due to Cameron 
and Martin (see, e.g., 
Refs.\cite{6,13,4,klep} for contributions and references therein).  An
alternative approach is based on the path integrals formulation for
quantum mechanics, which represents a powerful tool to analyse these
problems using methods more familiar to physicists \cite{16,17}:
here, the problem appears as the computation of the partition function
of a quantum-harmonic oscillator with time dependent
frequency. Various quadratic functionals of Brownian motion have been
studied in details by physicists \cite{18} with a variety of
methods. They arise in a plethora of physical contexts, from polymers
in elongational flows \cite{19} to
Casimir/van der Waals interactions and general fluctuation induced
interactions \cite{20,21,22,23,24} where, in the language of the
harmonic oscillator, both the frequency and mass depend on time. Quadratic
functionals of Brownian motion also arise in the theory of
electrolytes when one computes the one-loop or fluctuation corrections
to the mean field Poisson-Boltzmann theory \cite{25,26,27,28}. Finally
we mention that functionals of Brownian motion also turn out to have
applications in computer science \cite{29}.

In order to calculate $G(\sigma)$ in Eq.~(\ref{G}) we resort to
the path integrals formulation.  Following Refs.\cite{boyer,boyer1}, we
introduce an auxiliary functional
 \vspace{1pc} 
\begin{equation}
\label{fc}
\Psi(x,t) = \mathbb{E}^x_t\left\{ \exp\left( - \frac{\sigma A_{\alpha}}{T} \int^{T}_{t} \, \omega(\tau) \, B_{\tau}^2  \, d\tau   \right)\right\}
\end{equation}\vspace{1pc}  
where the expectation is for a Brownian motion starting at $x$ at time
$t$.  In terms of $\Psi(x,t)$ the moment-generating function is
determined by noting that $G(\sigma) = \Psi(0,0)$.

Further on, we
derive a Feynman-Kac type formula for $\Psi(x,t)$ considering how the
functional in Eq.~(\ref{fc}) evolves in the time interval
$(t,t+dt)$. During this interval the Brownian motion moves from $x$ to $x +
dB_t$, where $dB_t$ is an infinitesimal Brownian increment such
that $\mathbb{E}_{dB}\{dB_t\} = 0$ and $\mathbb{E}_{dB}\{dB^2_t\} = 2 D dt$, where
$\mathbb{E}_{dB}$ denotes now averaging with respect to the increment
$dB_t$. For such an evolution we have, to linear order in $dt$
\begin{eqnarray}
\fl
\Psi(x,t) = \mathbb{E}_{dB}\Big\{ \left(1 - \frac{\sigma A_{\alpha} \omega(t)}{T} x^2 dt\right)
\,
\mathbb{E}^{x+dB_t}_{t+dt}\left\{ \exp\left( - \frac{\sigma A_{\alpha}}{T} \int^{T}_{t+dt}  \omega(\tau) \, B_{\tau}^2 \, d\tau \right)\right\}  \Big\} \nonumber\\= \mathbb{E}_{dB}\left\{ \Psi(x + dB_t,t + dt) \left(1 - \frac{\sigma A_{\alpha} \omega(t)}{T} x^2 dt\right)\right\} \,.
\end{eqnarray}
Expanding the  right-hand-side of the latter equation  to second order
in  $dB_t$ and to linear  order in  $dt$,  we eventually find, after 
averaging, that      $\Psi(x,t)$      obeys     the      equation
\vspace{1pc} 
\begin{equation}
\label{17}
\frac{\partial \Psi(x,t)}{\partial t} = - D \frac{\partial^2 \Psi(x,t)}{\partial x^2} +
\frac{\sigma A_{\alpha} \omega(t)}{T} x^2 \Psi(x,t) \,,
\end{equation}\vspace{1pc}  
which looks like a Schr\"odinger equation with a harmonic
time-dependent potential. Eq.~(\ref{17}) is to be solved subject to
boundary condition $\Psi(x,T) = 1$ for any $x$.

We seek the solution of Eq.~(\ref{17}) for arbitrary $\omega(t)$ in the form
\vspace{1pc}  \begin{equation}
\Psi(x,t) = f(t) \exp\left(- \frac{1}{2} g(t) x^2\right) \,,
\end{equation}\vspace{1pc}  
where
\vspace{1pc}  \begin{equation}
\dot{f} = D f g \,\,, f(t=T) = 1 \,,
\end{equation}\vspace{1pc}  
and
\vspace{1pc}  \begin{equation}
\label{g}
\dot{g} = 2 D g^2 - \frac{\sigma \omega(t)}{d D \int^T_0 \tau \omega(\tau) d\tau} \ , \quad  g(t=T) = 0\,.
\end{equation}\vspace{1pc}  
Next, we get rid of the nonlinearity in Eq.(\ref{g}) introducing a new function
$h$ obeying
\vspace{1pc}  \begin{equation}
g = - \frac{1}{2 D} \frac{\dot{h}}{h} \,.
\end{equation}\vspace{1pc}  
The function $h(t)$ is solution of the linear second-order differential equation
\vspace{1pc}  \begin{equation}
\label{22}
\ddot{h} - \frac{2 \sigma \omega(t)}{d \int^T_0 \tau \omega(\tau) d\tau} h = 0 \,,
\end{equation}\vspace{1pc}  
which has to be solved subject to the boundary conditions
\vspace{1pc}  \begin{equation}
\label{bc}
h(t = T) = 1\, , \,\,\, \dot{h}(t = T) = 0 \,.
\end{equation}\vspace{1pc}  
Once $h(t)$ is found,  $f(t)$ is determined by $f(t) = 1/\sqrt{h(t)}$ and $G(\sigma)$ by $G(\sigma) = f(0) =1/\sqrt{h(t=0)}$.

\subsection{The moment-generating function for $\alpha \neq 2$.}

We focus now on
the particular case of the weight function
$\omega(t)$ defined by Eq.~(\ref{omega}) with $\alpha \neq 2$.
In this case Eq.~(\ref{22}) reads
\vspace{1pc}  \begin{equation}
\label{power}
\ddot{h} - \frac{a \sigma}{\left(t_0 + t\right)^{\alpha}} h = 0 \,,
\end{equation}\vspace{1pc}  
with
\vspace{1pc}  \begin{equation}
\label{a}
a = \frac{2 }{d \int^T_0 \tau (t_0 + \tau)^{-\alpha} d\tau} > 0 \,.
\end{equation}\vspace{1pc}  
Solution of Eq.(\ref{power}) has the form
\vspace{1pc}  \begin{equation}
\fl
h(t) = \sqrt{t_0 + t} \left[C_1 I_{\nu}\left(2 \nu \sqrt{a (t_0 + t)^{2 - \alpha} \sigma} \right) + C_2 K_{\nu}\left(2 \nu \sqrt{a (t_0 + t)^{2 - \alpha} \sigma} \right)\right] \,,
\end{equation}\vspace{1pc}  
where $I_{\mu}(\cdot)$ and $K_{\mu}(\cdot)$ are the modified Bessel functions \cite{abramowitz},
the exponent $\nu$ is given by
\vspace{1pc}  \begin{equation}
\label{nu}
\nu = \frac{1}{2 - \alpha} \,,
\end{equation}\vspace{1pc}  
while
the constants $C_1$ and $C_2$ are chosen to fulfil the boundary conditions in  Eqs.(\ref{bc}), so that
\vspace{1pc}  \begin{equation}
C_1 = 2 \nu \sqrt{a (t_0 + T)^{1 - \alpha} \sigma} K_{\nu - 1}\left(2 \nu \sqrt{a (t_0 + T)^{2 - \alpha} \sigma}\right) \,,
\end{equation}\vspace{1pc}  
and
\vspace{1pc}  \begin{equation}
C_2 = 2 \nu \sqrt{a (t_0 + T)^{1 - \alpha} \sigma} I_{\nu - 1}\left(2 \nu  \sqrt{a (t_0 + T)^{2 - \alpha} \sigma}\right) \,.
\end{equation}\vspace{1pc}  
Consequently, we find that $h(t=0)$ obeys
\begin{eqnarray}
\label{mm}
h(t=0)&=&
 \left(\frac{\epsilon}{1 + \epsilon}\right)^{(\alpha-1)/2} \frac{\pi \nu z \sqrt{\sigma}}{2 \sin\left(\pi \nu\right)} \times \nonumber\\
  &\times& \Big[I_{-\nu}\left(\nu z \sqrt{\sigma}\right) I_{\nu - 1}\left(\left(\frac{1 + \epsilon}{\epsilon}\right)^{1 - \alpha/2} \nu z \sqrt{\sigma}\right) - \nonumber\\
&-& I_{\nu}\left(\nu z \sqrt{\sigma}\right) I_{1-\nu}\left(\left(\frac{1 + \epsilon}{\epsilon}\right)^{1 - \alpha/2} \nu z \sqrt{\sigma}\right)\Big] \,,
\end{eqnarray}
where $\epsilon = t_0/T$,
 and
\vspace{1pc}  \begin{equation}
z = \sqrt{\frac{8 (1 - \alpha) (2 - \alpha) \epsilon^{2 - \alpha}}{d \left(\epsilon^{2 - \alpha} - (\alpha + \epsilon - 1) (1 + \epsilon)^{1 - \alpha} \right)}} \,.
\end{equation}\vspace{1pc}  
Turning finally
 to the limit $\epsilon \to 0$, we find that
 the leading small-$\epsilon$ behaviour of the moment-generating function
is given by
\vspace{1pc}  \begin{equation}
\label{alpha<2}
\Phi(\sigma) = \left[\Gamma\left(\nu\right) \left(\frac{\sigma}{\chi_1}\right)^{\frac{1-\nu}{2}} {\rm I}_{\nu - 1}\left(2 \sqrt{\frac{\sigma}{\chi_1}}\right)\right]^{-d/2} \,,
\end{equation}\vspace{1pc}  
for $\alpha < 2$, while for $\alpha > 2$ it obeys
\vspace{1pc}  \begin{equation}
\label{alpha>2}
\Phi(\sigma) = \left[\Gamma\left(1-\nu\right) \left(\frac{\sigma}{\chi_2}\right)^{\frac{\nu}{2}} {\rm I}_{-\nu}\left(2 \sqrt{\frac{\sigma}{\chi_2}}\right)\right]^{-d/2} \,,
\end{equation}\vspace{1pc}  
where
\vspace{1pc}  \begin{equation}
\label{chi}
\chi_1 = \frac{d (2 - \alpha)}{2} \, \,\,\,{\rm and} \,\,\, \chi_2 = \frac{d (\alpha - 2)}{2 (\alpha - 1)} \,.
\end{equation}\vspace{1pc}

\subsection{The moment-generating function for $\alpha = 2$.}

We  focus next  on the  particular case  $\alpha =  2$ and  consider 
for convenience a
slightly  more general  form of the  weight function  $\omega(t)$ by
introducing   an  additional  parameter   $\xi$.  We   stipulate  that
$\omega(t)$  is  the  piece-wise continuous  function
\vspace{1pc} \begin{equation}
\label{omegaalpha2}
\omega(t) =
\cases{2 \xi/t_0^2, & for $0 \leq t \leq t_0$,\\
1/t^2, & for $t_0 \leq t \leq T$.}
\end{equation}\vspace{1pc}  
We seek now  the corresponding  moment-generating
function  $\Phi(\sigma)$   and   optimise  it   in  what  follows
considering $\xi$ as an optimisation parameter.

The differential Eq.~(\ref{22}) has to be solved now for two intervals
$t \in [0,t_0]$  and $t \in [t_0,T]$ in which  the "potential" has two
different forms. For  the first interval, i.e., when  $t \in [0,t_0]$,
the      general       solution      of      Eq.~(\ref{22})      obeys
\vspace{1pc} \begin{equation} h(t) =  c_1 \, \exp\left(- \sqrt{2 a \xi
      \sigma} \, \frac{t}{t_0}\right) + c_2 \, \exp\left(\sqrt{2 a \xi
      \sigma} \, \frac{t}{t_0}\right) \,,
\end{equation}\vspace{1pc}  
where $c_1$ and $c_2$ are coefficients to be determined. The parameter $a$
given by Eq.~(\ref{a}) now reads
\vspace{1pc}  \begin{equation}
a= \frac{2}{d (\xi + \ln\left(1/\epsilon\right))} \,.
\end{equation}\vspace{1pc}  
For the second interval, i.e., when $t$ belongs to $[t_0, T]$, we have
\vspace{1pc}  \begin{equation}
h(t) = c_3 \, t^{(1 + \delta)/2} + c_4 \, t^{(1 - \delta)/2} \,,
\end{equation}\vspace{1pc}  
where
\vspace{1pc}  \begin{equation}
\label{delta}
\delta = \sqrt{1 + 4 a \sigma}\,,
\end{equation}\vspace{1pc}  
while the coefficients $c_3$ and $c_4$ are to be found from the boundary conditions in Eqs.~(\ref{bc}). This yields
\vspace{1pc}  \begin{equation}
c_3 = \frac{\delta - 1}{2 \delta} \, T^{-(1+\delta)/2} \,,
\end{equation}\vspace{1pc}  
and
\vspace{1pc}  \begin{equation}
c_4 = \frac{\delta + 1}{2 \delta} \, T^{(\delta - 1)/2} \,.
\end{equation}\vspace{1pc}  
Further on, we require the continuity of $h(t)$ and its first derivative at $t = t_0$. We find then that
\vspace{1pc}  \begin{equation}
c_1 = \frac{(\delta + 1) \exp\left(\sqrt{2 a \xi \sigma}\right)}{4 \delta \epsilon^{(\delta-1)/2}} \, \left( 1 + \frac{\delta -1}{\delta+1} \epsilon^{\delta} - \frac{\delta - 1}{2 \sqrt{2 a \xi \sigma}} \left(1 + \epsilon^{\delta}\right)\right)
\end{equation}\vspace{1pc}  
and
\vspace{1pc}  \begin{equation}
c_2 = \frac{(\delta + 1) \exp\left(-\sqrt{2 a \xi \sigma}\right)}{4 \delta \epsilon^{(\delta-1)/2}} \, \left( 1 + \frac{\delta -1}{\delta+1} \epsilon^{\delta} + \frac{\delta - 1}{2 \sqrt{2 a \xi \sigma}} \left(1 + \epsilon^{\delta}\right)\right)\,.
\end{equation}\vspace{1pc}  
Consequently, we find that in this case the moment-generating function is given for arbitrary $\epsilon$ explicitly by
\begin{eqnarray}
\label{alpha=2}
\Phi(\sigma) = \left(c_1 + c_2\right)^{-d/2} &=& \Big[\frac{(\delta + 1)}{2 \delta \epsilon^{(\delta-1)/2}} \Big(\Big( 1 + \frac{\delta -1}{\delta+1} \epsilon^{\delta}\Big) \, \cosh\left(\sqrt{2 a \xi \sigma}\right) + \nonumber\\ &+& \frac{\delta - 1}{2 \sqrt{2 a \xi \sigma}} \left(1 + \epsilon^{\delta}\right) \, \sinh\left(\sqrt{2 a \xi \sigma}\right)\Big) \Big]^{-d/2} \,.
\end{eqnarray}
Now, we are equipped with all necessary results to determine the variance of
the distribution $P(u_{\alpha})$ as well as the distribution itself.

\section{The variance of the distribution $P(u_{\alpha})$.}
\label{variance}

In  this  section we  analyse  the  behaviour  of the  variance  ${\rm
  Var}(u_{\alpha})$  of  the estimator  in  Eq.~(\ref{u}).  First,  we
calculate  exactly   the  limiting  small-$\epsilon$   form  of  ${\rm
  Var}(u_{\alpha})$  for arbitrary  $\alpha \neq  2$.  Further  on, we
focus on  the case $\alpha =  2$ and determine  ${\rm Var}(u_{\alpha =
  2})$ for arbitrary $\epsilon$ and $\xi$, Eq.~(\ref{omegaalpha2}). We
show next that the variance has a minimum for a certain amplitude $\xi
=  \xi_{\rm  opt}$ and  present  a  corresponding  expression for  the
optimised   variance.  Finally,   we  consider   the  case   when  the
continuous-space  and  -time   trajectory  is  recorded  at  arbitrary
discrete time moments $t_j$  and calculate exactly the weight function
$\omega(t)$ which provides the minimal possible variance.

\subsection{The variance for $\alpha \neq 2$ and $\epsilon = 0$.}

\begin{figure}[ht]
  \centerline{\includegraphics*[width=0.75\textwidth]{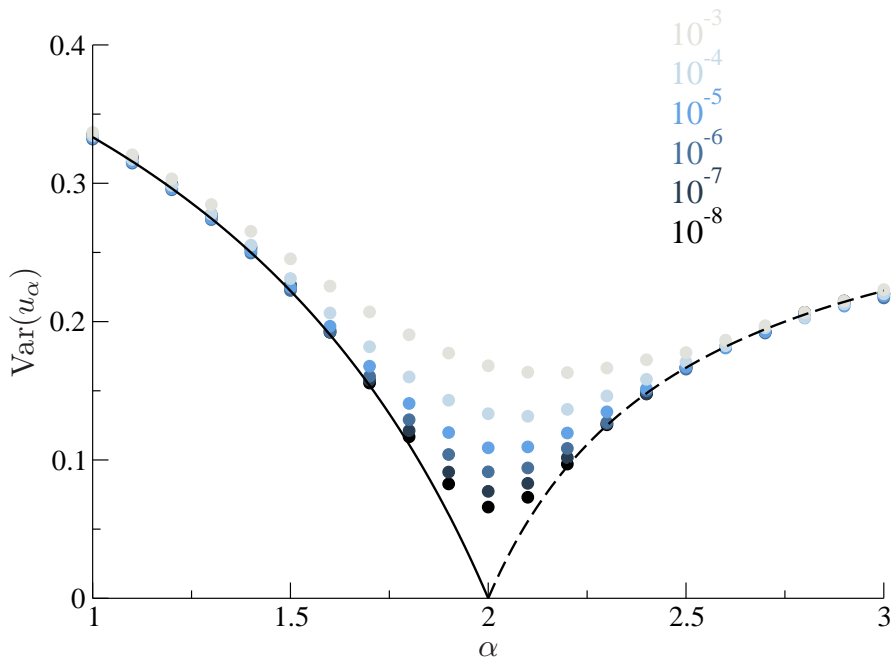}}
  \caption{(Color   online)   The   variance   of   the   distribution
    $P(u_{\alpha})$ in $d=3$ for different  values of  $\alpha$. Solid  line -
    Eq.~(\ref{varvar})  with  $\alpha  <  2$  and the  dashed  line  -
    Eq.~(\ref{varvar}) with  $\alpha > 2$.  The  symbols correspond to
    the values obtained from  numerical simulations of 3D random walks
    for   different  $\epsilon$, as indicated by the labels.}
\label{fig1}
\end{figure}

The variance ${\rm Var}(u_{\alpha})$ is obtained by differentiating
Eqs.~(\ref{alpha<2}) or (\ref{alpha>2}) twice with respect to $\sigma$
and setting $\sigma$ equal to zero.  For arbitrary $\alpha \neq 2$ the
variance is then given explicitly by \vspace{1pc} \begin{equation}
\label{varvar}
{\rm Var}(u_{\alpha}) = \frac{2}{d}
\cases{\frac{2 - \alpha}{3 - \alpha}, & for $\alpha < 2$,\\
\frac{\alpha - 2}{2 \alpha - 3}, & for $\alpha > 2$.}
\end{equation}\vspace{1pc}  
The result in  the latter equation is depicted  in Fig.~\ref{fig1} and
shows that, strikingly, the variance  can be made arbitrarily small in
the leading order in $\epsilon$ by taking $\alpha$ gradually closer to
$2$, either from above or from below. The slopes at $\alpha = 2^+$ and
$\alpha =  2^-$ appear to be the  same, so that the  accuracy of the
estimator will be the same for  approaching $\alpha = 2$ from above or
from below. Equation (\ref{varvar}),  although formally invalid for $\alpha
= 2$, also suggests that the estimator in Eq.~(\ref{u}) with $\alpha =
2$ has vanishing variance and thus possesses an ergodic property.

A word of caution is now in order. As a matter of fact, we deduce from
Eq.~(\ref{mm}) that finite-$\epsilon$ corrections to the result in
Eq.~(\ref{varvar}) are of order of $\mathcal{O}(\epsilon^{2 -
  \alpha})$ for $1 < \alpha < 2$, which means that this asymptotical
behaviour can be only attained when $\epsilon \ll \exp\left(-1/(2 -
  \alpha)\right)$.  In other words, the variance can be made
arbitrarily small by choosing $\alpha$ closer to $2$, but only at
expense of either decreasing the experimental resolution time $t_0$ or 
increasing the
observation time $T$, which is clearly seen from the numerical
data shown in Fig.~\ref{fig1}.

\subsection{The variance for $\alpha = 2$ and arbitrary $\epsilon$.}
\label{arbitrary}

Differentiating Eq.~(\ref{alpha=2}) with respect to $\sigma$ twice, we find
that for arbitrary $\epsilon$
the variance of the distribution $P(u_{2})$
is given explicitly by
\vspace{1pc}  \begin{equation}
\label{7}
{\rm Var}(u_2) = \frac{4}{3 d} \, \frac{3 \ln(1/\epsilon) - 3 (1-\epsilon) + 2 (1 - \epsilon) \xi + \xi^2}{\left(\xi + \ln(1/\epsilon)\right)^2} \,.
\end{equation}\vspace{1pc}  
Notice now that ${\rm Var}(u_2)$ in Eq.~(\ref{7})
is a non-monotonic function of $\xi$
which attains
its minimal value when
\vspace{1pc}  \begin{equation}
\label{xi}
\xi = \xi_{{\rm opt}} = \frac{(2 + \epsilon) \ln(1/\epsilon) - 3 (1 - \epsilon)}{\ln(1/\epsilon) + \epsilon - 1} \,.
\end{equation}\vspace{1pc}  
This is a somewhat surprising result which states that the optimal
choice of the amplitude $\xi$ in Eq.~(\ref{omegaalpha2}), which
defines the weight function $\omega(t)$, actually depends on both the
time lag $t_0$ and on the observation time $T$.  In other words, in
order to make a proper choice of the amplitude $\xi$, one has to know
in advance the time through which the trajectory is observed.  Similar
dependence of the optimal parameters on the observation time has been
recently reported in Refs.\cite{katja1,katja2}, which optimised the
number of distinct sites visited by intermittent random walks. Note
that $\xi_{\rm opt} \to 2$ as $\epsilon \to 0$, but for any finite
$\epsilon$ it is greater than $2$.

Plugging the  expression in  Eq.~(\ref{xi}) into the  Eq.~(\ref{7}) we
obtain        the       corresponding        optimised       variance
\vspace{1pc} \begin{equation}
\label{optimal}
{\rm Var}_{\rm opt}(u_2) = \frac{4 }{3 d} \, \frac{3 \ln(1/\epsilon) - 4 + 5 \epsilon - \epsilon^2}{\ln(1/\epsilon) \left(\ln(1/\epsilon) + 1 + 2 \epsilon\right) - 3 (1 - \epsilon)} \,.
\end{equation}\vspace{1pc}  
From Eq.~(\ref{optimal}) we find  that in 3D ${\rm Var}_{\rm opt}(u_2)
\approx  0.144$ for  $\epsilon =  10^{-3}$, ${\rm  Var}_{\rm opt}(u_2)
\approx 0.096$ for $\epsilon  = 10^{-5}$ and ${\rm Var}_{\rm opt}(u_2)
\approx 0.082$ for $\epsilon =  10^{-6}$. When $\epsilon \to 0$, ${\rm
  Var}_{\rm opt}(u_2)$  vanishes in a logarithmic way with $\epsilon$ at 
  leading order: \vspace{1pc} \begin{equation}
\label{asymptotic1}
{\rm Var}_{\rm opt}(u_2) \sim \frac{4}{d} \frac{1}{\ln(1/\epsilon)}\,.
\end{equation}\vspace{1pc}  
Therefore, the variance  can be made arbitrarily small  but at expense
of  a   progressively  higher  resolution  or   a  larger  observation
time. Note  that this  is the only  case ($\alpha  = 2$) in  which the
estimator defined by Eq.~(\ref{u}) is ergodic.

\subsection{Optimisation of the variance for continuous-time and -space trajectories recorded at discrete time moments.}
\label{discretet}

Finally  we consider the  estimation of  the diffusion  constant $D_f$
when one has  a set of $N$ temporal points  $t_j$ such that $0<t_0<t_1
<\cdots <  t_{N-1}=T$ and  a value $B_{t_j}^2$,  (which is one  of the
components  of $d$-dimensional  Brownian  motion), for  each of  these
points.       We     consider     the      least-squares     estimator
\vspace{1pc} \begin{equation}
\label{diss}
u_{\rm dis} = {1\over 2} \sum_{j=0}^{N-1} \omega_j \, B^2_{t_j} \,,
\end{equation}\vspace{1pc}  
where the normalisation is now adsorbed into
the weight function $\omega_j$, so that
\vspace{1pc}  \begin{equation}
{1\over 2} \sum_{j=0}^{N - 1} \omega_j \, \mathbb{E}\{B^2_{t_j}\} = 1\label{condis}
\end{equation}\vspace{1pc}  
As  in the  previous subsection,  we seek  an optimal  weight function
$\omega_j$ which minimises the variance of the least-squares estimator
in   Eq.~(\ref{diss}).  Remarkably,   this  problem   can   be  solved
\textit{exactly} for any arbitrary set $\{t_j\}$.

The  variance of  this estimator  can be  straightforwardly calculated
explicitly,      for      arbitrary      $\omega_j$,      to      give
\vspace{1pc}  \begin{equation} {\rm Var}(u_{\rm  dis}) =  2 \sum_{j,k}
  \omega_j \, \omega_k \, (t_j \wedge t_k)^2 \,,
\end{equation}\vspace{1pc}  
where $(t_j \wedge t_k)$ equals  the smallest of two numbers $t_j$ and
$t_k$.

In order to determine the choice of the $\omega_j$ which minimises the
variance    of   the   estimator,    we   minimise    the   functional
\vspace{1pc} \begin{equation}  F =  {1\over 2} \sum_{j,k}  \omega_j \,
  (t_j  \wedge  t_k)^2 \,  \omega_k  - \lambda  \left(\sum_{j=0}^{N-1}
    \omega_j t_j -1\right) \,,
\end{equation}\vspace{1pc}  
where    $\lambda$     is    a    Lagrange     multiplier    enforcing
Eq.~({\ref{condis}).  Minimising with respect to each $\omega_j$ gives
  the  system   of  linear  equations   \vspace{1pc}  \begin{equation}
    \sum_{j}    (t_j\wedge    t_k)^2    \omega_j    =    \lambda    \,
    t_k.\label{optdis}
\end{equation}\vspace{1pc}  
To solve this system of equations exactly, we define
\vspace{1pc}  \begin{equation}
\Omega_k = \sum_{j \geq k} \omega_j \,,
\end{equation}\vspace{1pc}  
or, equivalently,
\vspace{1pc}  \begin{equation}
\omega_j = \Omega_j -\Omega_{j+1} \,,
\end{equation}\vspace{1pc}  
for  $0 \leq j\leq  N-2$. Also  clearly we  have that  $\Omega_{N-1} =
\omega_{N-1}$  which  is   compatible  with  defining  $\Omega_{N}=0$.
Therefore      Eq.~(\ref{optdis})      can      be     written      as
\vspace{1pc} \begin{equation} \sum_{j<k} (\Omega_j-\Omega_{j+1}) t_j^2
  + t_k^2 \Omega_ k = \lambda t_k \label{Omega}.
\end{equation}\vspace{1pc}  
Now subtracting  Eq.~(\ref{Omega}) for $k$ from the  same equation for
$k+1$ gives the  solution \vspace{1pc} \begin{equation} \Omega_{k+1} =
  {\lambda\over t_{k+1} + t_{k}},
\end{equation}\vspace{1pc}  
valid for $0 \leq k\leq N-2$, which implies that
\vspace{1pc}  \begin{equation}
\Omega_{k} = {\lambda\over t_{k} + t_{k-1}}\label{int}
\end{equation}\vspace{1pc}  
which is valid for $1\leq k\leq  N-1$. In addition, if we set $k=0$ in
Eq. (\ref{Omega}) we  find that \vspace{1pc} \begin{equation} \Omega_0
  = {\lambda \over t_0},
\end{equation}\vspace{1pc}  
which is  compatible with Eq.~(\ref{int}) upon  defining an additional
point  $t_{-1}  =0$.  We  thus   find  that  for  $1\leq  j\leq  N-2$,
\vspace{1pc} \begin{equation}
\label{k1}
\omega_j = \Omega_j - \Omega_{j+1} = {\lambda \over (t_{j}+t_{j-1})} -  {\lambda \over (t_{j}+t_{j+1})}\,,
\end{equation}\vspace{1pc}  
while
\vspace{1pc}  \begin{equation}
\label{k2}
\omega_0 = {\lambda t_1\over t_0(t_0+t_1)} \,,
\end{equation}\vspace{1pc}  
and
\vspace{1pc}  \begin{equation}
\label{k3}
\omega_{N-1} = \Omega_{N-1} = {\lambda\over t_{N-1} + t_{N-2}} \,.
\end{equation}\vspace{1pc}  
 The normalisation constraint, Eq. (\ref{condis}), then yields $\lambda$ as
 \vspace{1pc}  \begin{equation}
 \label{k4}
 \lambda = \left( {t_1\over t_0+t_1} + {t_{N-1}\over t_{N-1} + t_{N-2}} + \sum_{j=1}^{N-2} {t_j(t_{j+1}-t_{j-1})
 \over (t_{j+1}+ t_j)(t_j+t_{j-1})}\right)^{-1}.
 \end{equation}\vspace{1pc}  
 Finally the minimal variance can be computed by multiplying Eq.(\ref{optdis})
 by    $\omega_k$     and    summing    over     $k$    which    gives
 \vspace{1pc}  \begin{equation}   \sum_{j,k}  \omega_j  \,  (t_j\wedge
   t_k)^2  \omega_k = \lambda  \, \sum_{j=0}^{N-1}  \omega_j \,  t_j =
   \lambda \,, \label{optdis2}
\end{equation}\vspace{1pc}  
and hence,
\vspace{1pc}  \begin{equation}
{\rm Var}(u_{\rm dis}) = 2 \sum_{j,k} \omega_j \, \omega_k (t_j\wedge t_k)^2 = 2 \, \lambda \,,
\end{equation}\vspace{1pc}  
where    we   have    again   used    the    normalisation   condition
Eq.~(\ref{condis}).  Equations  (\ref{k1})  to (\ref{k4})  define  the
exact solution  of the problem  of the optimal estimator  for Brownian
trajectories recorded at discrete time moments.

In order  to compare the results  with the continuum case  we take the
first point $t_0$ to be fixed and write $t_j = t_0 + \Delta (j-1)$ for
$j >  0$, where $\Delta$ is  a constant time step,  $\Delta = T/(N -  1)$. 
This gives  the following  expression  for the  Lagrange multiplier,  which
defines       the       variance       of      the       distribution,
\vspace{1pc} \begin{equation}  \fl \lambda  = \left({t_0\over 2  t_0 +
      \Delta}   +   {T\over   2   T   -  \Delta}   +   2   \Delta   \,
    \sum_{j=1}^{N-2}{t_0  +   (j-1)  \Delta   \over  (2t_0  +   (2j  -
      1)\Delta)(2 t_0 + (2j +1) \Delta)} \right)^{-1} \,.
 \end{equation}\vspace{1pc}  
 Turning to  the limit $\Delta\to 0$  and $N \to  \infty$, but keeping
 the  ratio $T  =  N/\Delta$ fixed,  the  sum in  the latter  equation
 becomes a Riemann integral  and we find \vspace{1pc} \begin{equation}
   \lambda^{-1}  = 1 +  {1\over 2}  \int_{t_{0}}^T {dt\over  t} =  1 +
   {1\over 2}\ln({T\over t_0}) \,,
 \end{equation}\vspace{1pc}  
so that in the leading in $\epsilon$ order,
for $d$-dimensional systems,
 \vspace{1pc}  \begin{equation}
 \label{discrete}
{\rm Var}(u_{\rm dis}) = {4\over d \left(2 + \ln({1/\epsilon})\right)} \,.
 \end{equation}\vspace{1pc}  
 Note that  for $\ln(1/\epsilon) \gg 2$, the  latter equation exhibits
 exactly the same asymptotic behaviour  in the limit $\epsilon \to 0$,
 as   the   result    of   the   previous   Section   \ref{arbitrary},
 Eq.~(\ref{asymptotic1}).

\section{Tails of the distribution $P(u_{\alpha})$ in $d$ dimensions}
\label{asymp}

Exact  expressions  for the  moment-generating  function  allow us  to
deduce  the asymptotic behaviour  of the  distribution $P(u_{\alpha})$
for $u_{\alpha} \ll 1$ and $u_{\alpha} \gg 1$.

\subsection{Asymptotic behaviour of $P(u_{\alpha})$ for $\alpha \neq 2$ and $\epsilon = 0$.}

Large-  and small-$u_{\alpha}$ asymptotics  of $P(u_{\alpha})$  can be
deduced directly  from Eqs.~(\ref{alpha<2}) and  (\ref{alpha>2}).  Let
us first  focus on the small-$u_{\alpha}$ behaviour,  which stems from
the  large-$\sigma$ asymptotical  behaviour  of the  moment-generating
function.     For     $\alpha     <     2$    the     latter     obeys
\vspace{1pc} \begin{equation}
\label{largesigma}
\Phi(\sigma) \sim \sigma^{d (1 + 2/(2 - \alpha))/8} \, \exp\left(- \sqrt{\frac{2 d \sigma}{2 - \alpha}}\right) \,.
\end{equation}\vspace{1pc}  
Inverting Eq.~(\ref{largesigma})  we find that for  $u_{\alpha} \to 0$
and  $\alpha < 2$  the distribution  $P(u_{\alpha})$ shows  a singular
behaviour: \vspace{1pc} \begin{equation}
\label{short}
P(u_{\alpha}) \sim \exp\left(- \frac{d^2}{4 \chi_1 } \cdot \frac{1}{u_{\alpha}}\right) \frac{1}{u_{\alpha}^{\zeta}} \,,
\end{equation}\vspace{1pc}  
where the exponent $\zeta$ is given by
\vspace{1pc}  \begin{equation}
\zeta = \frac{3}{2} + \frac{d}{4} \frac{\alpha}{|2 - \alpha|}\,,
\end{equation}\vspace{1pc}  
and  the  parameter  $\chi_1$  is  defined  in  Eq.~(\ref{chi}).   The
analogous  left  tail for  $\alpha  > 2$  case  can  be obtained  from
Eq.~(\ref{short})  by  simply   making  the  replacement  $\chi_1  \to
\chi_2$.

Note that Eq.~(\ref{short}) describes  a bell-shaped function whose
maximum gives an approximation to the
most  probable  value of the estimator $u$ \vspace{1pc}  \begin{equation}  u^*  =  \frac{2
    d}{\alpha d + 6 |2 - \alpha|}
\end{equation}\vspace{1pc}  
Note that for any fixed $d$ and $\alpha \to 2$, the most probable $u^*
\to  1$,  i.e. to  the  ensemble average  value  of  the estimator  in
Eq.~(\ref{u}). Therefore, the  least-squares estimators outperform the
naive  end-to-end  estimator   of  the  diffusion  coefficient,  whose
distribution is given in Eq.  (\ref{chi-s}) and has a bell-shaped form
only for $d \geq 3$.

Next, we turn to  the large-$u_{\alpha}$ asymptotical behaviour of the
distribution function.  To do  this, it is  convenient to  rewrite the
result in Eq.~(\ref{alpha<2}) as \vspace{1pc} \begin{equation}
\label{product}
\Phi(\sigma) = \prod_{m = 1}^{\infty} \left(1 + 8 \sigma/(2 - \alpha) d \gamma^2_{\nu-1,m}\right)^{-d/2} \,,
\end{equation}\vspace{1pc}  
where  $\gamma_{\mu,m}$ is  the  $m$-th zero  of  the Bessel  function
$J_{\mu}(z)$, organised in  an ascending order \cite{abramowitz}.  The
large-$u_{\alpha}$ behaviour  of the distribution  function stems from
the behaviour of the moment-generating function in the vicinity of the
singular point on the negative  $\sigma$-axis which is
closest to  $\sigma = 0$ (all singularities  are  all located on
the  negative  $\sigma$-axis).   This  yields,  for $\alpha  <  2$,  an
exponential decay of the form \vspace{1pc} \begin{equation}
\label{long1}
P(u_{\alpha}) \sim u_{\alpha}^{d/2 - 1} \exp\left(- \frac{\chi_1 \gamma_{\nu-1,1}^2}{4} \cdot u_{\alpha}\right) \,.
\end{equation}\vspace{1pc}  
In   a  similar   fashion,   we   get  that   for   $\alpha  >2$   the
moment-generating      function     can     be      represented     as
\vspace{1pc} \begin{equation}
\label{product2}
\Phi(\sigma) = \prod_{m = 1}^{\infty} \left(1 + 8 (\alpha - 1) \sigma/(\alpha - 2) d \gamma^2_{-\nu,m}\right)^{-d/2} \,,
\end{equation}\vspace{1pc}  
so that
in this case the right tail of $P(u_{\alpha})$ follows
\vspace{1pc}  \begin{equation}
\label{long2}
P(u_{\alpha}) \sim u_{\alpha}^{d/2 - 1} \exp\left(- \frac{\chi_2 \gamma_{-\nu,1}^2}{4} \cdot u_{\alpha}\right) \,.
\end{equation}\vspace{1pc}  

To summarise the results of this subsection, we note the following:
\begin{itemize}

\item  when $\alpha  \to  2$, either  from  above or  from below,  the
  small-$u_{\alpha}$ behaviour of $P(u_{\alpha})$ becomes progressively more singular and
  small  values of $u_{\alpha}$  become very unlikely
  since  both $\chi_1$  and $\chi_2$  tend  to zero  and the  exponent
  $\zeta$ diverges.

\item  when $\alpha  \to  2$, either  from  above or  from below,  the
  inverse  relaxation  "lengths"  (i.e.,   the  terms  $(\alpha  -  2)
  \gamma_{-\nu,1}^2$  and  $(2 -  \alpha)  \gamma_{\nu-1,1}^2$ in  the
  exponentials  in   Eqs.~(\ref{long1})  and  (\ref{long2}))  diverge,
  suppressing large values of $u_{\alpha}$ in the distribution.

\end{itemize}

Since $P(u_{\alpha})$  is normalised  for arbitrary $\alpha$,  so that
the  area  below  the  curve  is  \textit{fixed},  this  implies  that
$P(u_{\alpha})$ tends to the delta-function as $\alpha \to 2$.

\subsection{Asymptotic behaviour of $P(u_{\alpha})$ for $\alpha = 2$ and small $\epsilon$.}

We focus first on the left  tails of the distribution for $\alpha = 2$
and fixed  small $\epsilon$. From Eq.~(\ref{alpha=2}) we  find that in
the limit  $\sigma \to \infty$ (so that  $\delta$ in Eq.~(\ref{delta})
is  $\delta   \gg  1$),  fixed  sufficiently   small  $\epsilon$,  the
moment-generating   function   obeys   \vspace{1pc}   \begin{equation}
  \Phi(\sigma)  \sim \exp\left(- \sqrt{\frac{d  \, \ln(1/\epsilon)}{2}
      \cdot \sigma}\right) \,,
\end{equation}\vspace{1pc}  
from which equation we readily obtain the following singular form:
\vspace{1pc}  \begin{equation}
P(u_{2}) \sim \exp\left(- \frac{d \ln(1/\epsilon)}{8} \cdot \frac{1}{u_2}\right) \frac{1}{u_{2}^{3/2}} \,,
\end{equation}\vspace{1pc}  
which holds for $u_2 \ll 1$. Within the opposite limit, i.e., for $u_2
\gg1$, the leading behaviour of the distribution $P(u_2)$ is dominated
by   the  closest   to  the   origin  root   of  the   denominator  in
Eq. (\ref{alpha=2}). Some algebra gives  that for $\epsilon \to 0$ the
distribution   function  $P(u_2)$  has   the  following   simple  form
\vspace{1pc} \begin{equation}  P(u_{2}) \sim u_2^{d/2  - 1}\exp\left(-
    \frac{d \, x_0^2 \left( \xi_{\rm opt} + \ln(1/\epsilon) \right)}{4
      \, \xi_{\rm opt}} \cdot u_2\right) \,,
\end{equation}\vspace{1pc}    
where $\xi_{\rm opt}$ is the optimised amplitude in Eq. (\ref{xi}) and
$x_0$, in the limit $\epsilon \to 0$, is the root of the equation
\begin{equation}
\left(1 - \frac{2 x_0^2}{\xi_{\rm opt}} \right)^{1/2} \, \frac{x_0 \cos\left(x_0\right)}{\sin\left(x_0\right)} = \frac{1}{2 + \ln(1/\epsilon)}
\end{equation}
The asymptotic behaviour of $x_0$ can be readily obtained:
\begin{equation}
\fl
x_0 = \sqrt{\frac{\xi_{\rm opt}}{2}} \left(1 - \frac{1}{\xi_{\rm opt} \, \cot^2\left(\sqrt{\xi_{\rm opt}/2}\right) \, \left(2 + \ln(1/\epsilon)\right)^2} + \mathcal{O}\left(\frac{1}{\ln^4(1/\epsilon)}\right)\right) \,.
\end{equation}
Therefore,  the  characteristic  decay  lengths  of  the  distribution
$P(u_2)$   from    both   sides    from   the   maximum    vanish   as
$1/\ln(1/\epsilon)$ when $\epsilon \to 0$.

\section{The distribution $P(u_{\alpha})$ in $d$ dimensions}
\label{dist}

We  turn   now  to   the  inversion  of   the  Laplace   transform  in
Eq.~(\ref{laplace})   in   order   to   visualise   the full  distribution
$P(u_{\alpha})$ and  to get an idea  of the location  of most probable
values of the estimators in Eq.~(\ref{u}).

\subsection{Inversion of the Laplace transform for $\alpha \neq 2$}

As  we  have already  remarked,  all  poles  of the  moment-generating
function $\Phi(\sigma)$ lie on the  complex plane on the negative real
$\sigma$-axis,  as can  be readily  seen from  the  representations in
Eqs.~(\ref{product}) and  (\ref{product2}). Setting then  $\gamma = 0$
in Eq.~(\ref{def:dist}), we find \vspace{1pc} \begin{equation}
\label{distr}
P(u_{\alpha}) = \frac{1}{\pi} \int^{\infty}_0 \frac{dz \, \cos\left(z u_{\alpha} - d \, \phi_{\alpha}(z)/2\right)}{\rho_{\alpha}^{d/4}(z)},
\end{equation}\vspace{1pc}  
where, for $\alpha < 2$,
\begin{eqnarray}
\rho_{\alpha}(z) = \Gamma^2\left(\nu\right) \left(\frac{\chi_1}{z}\right)^{\nu - 1} \, \left({\rm ber}_{\nu-1}^2\left(2 \, \sqrt{\frac{z}{\chi_1}}\right)+{\rm bei}_{\nu -1}^2\left(2 \, \sqrt{\frac{z}{\chi_1}}\right)\right) \,,
\end{eqnarray}
and the phase $\phi$ is given by
\vspace{1pc}  \begin{equation}
\phi_{\alpha}(z) = {\rm arctg}\left({\rm ber}_{\nu - 1}\left(2 \,\sqrt{\frac{ z}{\chi_1}}\right)/{\rm ber}_{\nu-1}\left(2 \, \sqrt{\frac{z}{\chi_1}}\right)\right) \,,
\end{equation}\vspace{1pc}  
while for $\alpha > 2$ we have
\begin{eqnarray}
\rho_{\alpha}(z) = \Gamma^2\left(1-\nu\right) \left(\frac{\chi_2}{z}\right)^{-\nu} \, \left({\rm ber}_{-\nu}^2\left(2 \, \sqrt{\frac{z}{\chi_2}}\right)+{\rm bei}_{-\nu}^2\left(2 \, \sqrt{\frac{z}{\chi_2}}\right)\right) \,,
\end{eqnarray}
and
\vspace{1pc}  \begin{equation}
\phi_{\alpha}(z) = {\rm arctg}\left({\rm ber}_{-\nu}\left(2 \, \sqrt{\frac{z}{\chi_2}}\right)/{\rm ber}_{-\nu}\left(2  \sqrt{\frac{z}{\chi_2}}\right)\right) \,,
\end{equation}\vspace{1pc}  
where  ${\rm ber}_{\mu}(x)$  and ${\rm  bei}_{\mu}(x)$ are  the Kelvin
functions   \cite{abramowitz}.  Equation  (\ref{distr})   defines  the
probability distributions $P(u_{\alpha})$ in the leading in $\epsilon$
order for  arbitrary $\alpha \neq  2$ and arbitrary  spatial dimension
$d$.

\begin{figure}[ht]
  \centerline{\includegraphics*[width=0.75\textwidth]{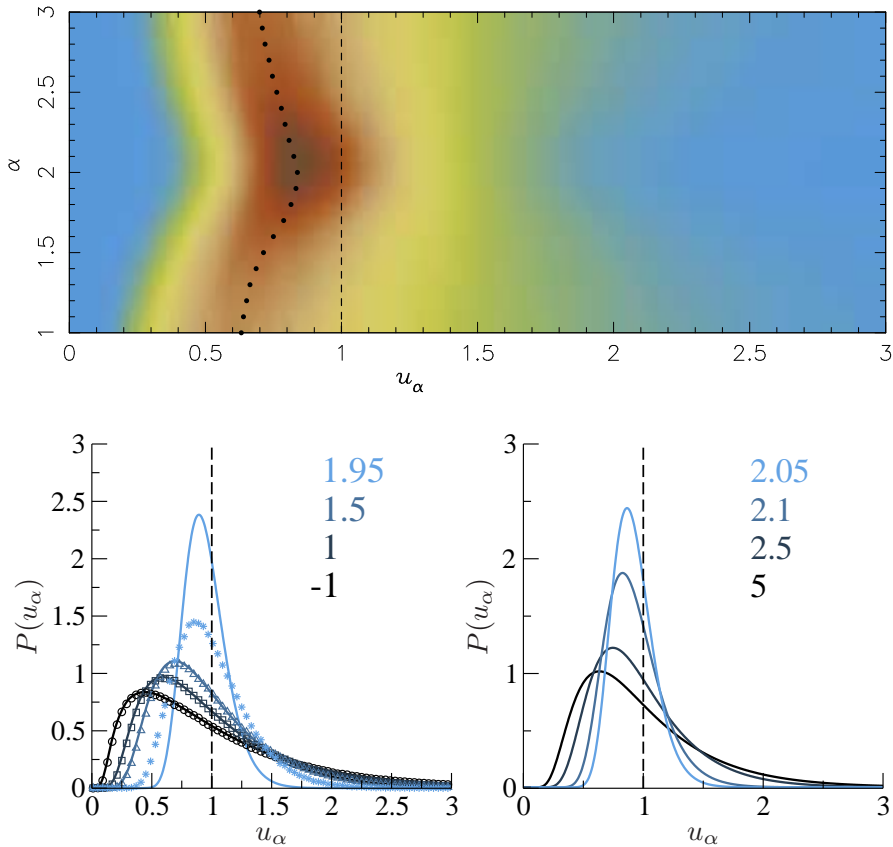}}
  \caption{(Color  online)  The  distribution  $P(u_{\alpha})$  in  3D
    systems. Upper  panel: Colour density map of  $P(u_{\alpha})$ as a
    function of  $\alpha$, obtained  from numerical simulations  of 3D
    random walks, with $\epsilon=10^{-5}$.  The solid knots indicate, for
    different values  of $\alpha$, the  position of the  most probable
    value   of   the  estimator   $u_{\alpha}$.    Lower  panel:   The
    distribution $P(u_{\alpha})$  for different $\alpha\ne2$ and with $\epsilon=0$, 
    obtained by numerical inversion of Eq.~(\ref{distr}) for $\alpha < 2$ (left
    lower  panel) and  of Eq.~(\ref{distr})  for $\alpha  >  2$ (right
    lower panel). The symbols in the left panel correspond to numerical
    simulations for  (from dark to  light), $\alpha = -  1$ (circles),
    $\alpha = 1$ (squares), $\alpha  = 3/2$ (triangles), and $\alpha =
    1.95$ (stars), and $\epsilon=10^{-5}$,  except for $\alpha = 1.95$
    for which we used $\epsilon=10^{-7}$.}
\label{fig2}
\end{figure}

In Fig. \ref{fig2} we plot $P(u_{\alpha})$ from Eq.~(\ref{distr}) for
$\alpha \neq 2$ and $\epsilon=0$ for three-dimensional systems together with the
results of numerical simulations. One notices that the theoretically
predicted distribution $P(u_{\alpha})$ becomes more narrow and its
maximal value grows when $\alpha$ moves towards $ \alpha = 2$, either
from above or from below. When $\alpha$ approaches $2$ from below, the
most probable value moves toward the ensemble average value ($=1$) of
the estimator and then starts to move back when $\alpha$ overpasses
$2$ and grows further.  Similarly to the behaviour of the variance, we
observe a very good agreement between our theoretical predictions and
numerical data for $\alpha $ not too close to $2$ for $\epsilon =
10^{-5}$ (lower left panel of Fig.\ref{fig2}). For $\alpha = 1.95$ and
$\epsilon$ as small as $10^{-7}$, we still see a discrepancy between
the numerical data and the asymptotic form of $P(u_{\alpha})$ in
Eq.~(\ref{distr}).  Note that the same slow convergence to zero
variance was observed in Fig.\ref{fig1} as $\epsilon\to0$.

\begin{figure}[ht]
  \centerline{\includegraphics*[width=0.75\textwidth]{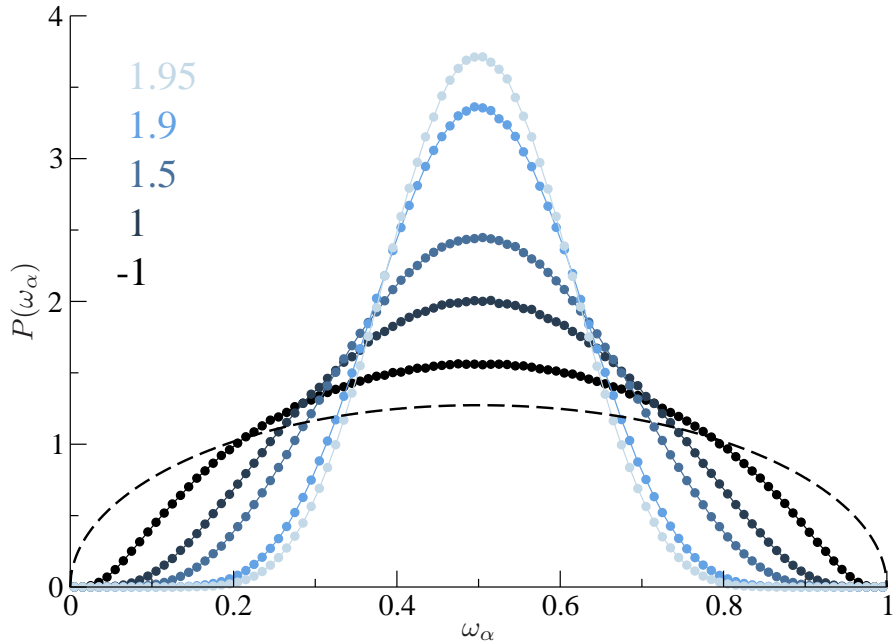}}
  \caption{(color  online)  The  distribution $P(\omega_{\alpha})$  in
    Eq.  (\ref{romega}) for  different  $\alpha <  2$  in 3D  systems.
    Symbols  are the  results  of numerical  simulations.  The  dashed
    curve is  the corresponding result for the  end-to-end estimate of
    the diffusion coefficient in Eq. (\ref{lim}).}
\label{fig3}
\end{figure}

In Fig.\ref{fig3} we present  the results of numerical simulations for
the distribution in Eq. (\ref{romega}) of the random variable $\omega$
defined in Eq.  (\ref{varomega}). One notices that as  $\alpha \to 2$,
the  distribution becomes progressively  narrower  and the  peak at
$\omega =  1/2$ becomes more  pronounced, which means that  it becomes
progressively  more likely  that the  diffusion  coefficients, deduced
from  two different  realisations of  Brownian trajectories  using the
weighted least-squares estimators, will have the same value.

\subsection{Two-dimensional systems: series representation of the distributions $P(u_{\alpha})$ and $P(\omega_{\alpha})$}

We proceed further  and focus now on the case $d =  2$ when $\Phi(\sigma)$
in Eqs.~(\ref{product})  and (\ref{product2}) have  only simple poles.
In  this   case,  we  readily  find  via   standard  residue  calculus
\vspace{1pc} \begin{equation}
\label{pu2d}
P(u_{\alpha}) = \frac{2^{-\nu}}{\Gamma(\nu + 1)} \, \sum_{m=1}^{\infty} \frac{\gamma^{\nu}_{\nu - 1,m}}{J_{\nu}\left(\gamma_{\nu-1,m}\right)} \, \exp\left( - \frac{(2 - \alpha) \gamma^2_{\nu-1,m}}{4} u_{\alpha}\right) \,,
\end{equation}\vspace{1pc}  
for $\alpha < 2$, while for $\alpha > 2$ we get
\vspace{1pc}  \begin{equation}
\label{pu2d1}
P(u_{\alpha}) = \frac{2^{\nu - 1}}{\Gamma(2 - \nu)} \, \sum_{m=1}^{\infty}
\frac{\gamma^{1-\nu}_{-\nu,m}}{J_{1-\nu}\left(\gamma_{-\nu,m}\right)} \,
\exp\left( - \frac{(\alpha - 2) \gamma^2_{-\nu,m}}{(\alpha - 1) 4} u_{\alpha}\right) \,.
\end{equation}\vspace{1pc}  
These expressions  can be readily  plotted using, e.g., Mathematica,
and they show  exactly the same qualitative behaviour (apart from a slightly larger
variance), as  our results  obtained by the  inversion of  the Laplace
transform for three-dimensional systems (see, Fig. \ref{fig2}).

We   finally   present   explicit   results   for   the   distribution
$P(\omega_{\alpha})$,    Eq.   (\ref{omega}),    for   two-dimensional
systems. For  $P(u_{\alpha})$ in Eq.  (\ref{pu2d}) one  has, using the
definition in Eq. (\ref{romega}), \vspace{1pc}
\begin{eqnarray}
\label{ss}
\fl
P(\omega_{\alpha}) &=& \left(\frac{2^{-\nu}}{\Gamma\left(\nu + 1\right) \, (1 - \omega_{\alpha})} \right)^2 \, \sum_{n=1}^{\infty} \frac{\gamma^{\nu}_{\nu - 1,n}}{J_{\nu}\left(\gamma_{\nu-1,n}\right)} \, \sum_{m=1}^{\infty} \frac{\gamma^{\nu}_{\nu - 1,m}}{J_{\nu}\left(\gamma_{\nu-1,m}\right)}  \nonumber\\
\fl
&\times& \int_0^{\infty} u_{\alpha} \, du_{\alpha} \, \exp\left(- \frac{(2 - \alpha)}{4} \, \left( \frac{\omega_{\alpha}}{1 - \omega_{\alpha}} \, \gamma^2_{\nu-1,n} +   \gamma^2_{\nu-1,m}\right) \, u_{\alpha}\right) \nonumber\\
\fl
&=& \left(\frac{2^{2-\nu}}{(2 - \alpha) \, \Gamma\left(\nu + 1\right) }\right)^2  \, \sum_{n=1}^{\infty} \frac{\gamma^{\nu}_{\nu - 1,n}}{J_{\nu}\left(\gamma_{\nu-1,n}\right)} \nonumber\\
\fl
&\times& 
 \sum_{m=1}^{\infty} \frac{\gamma^{\nu}_{\nu - 1,m}}{J_{\nu}\left(\gamma_{\nu-1,m}\right)} \,\frac{1}{\left(\omega_{\alpha}  \gamma^{2}_{\nu - 1,n} + (1- \omega_{\alpha}) \gamma^{2}_{\nu - 1,m}\right)^2} \,.
\end{eqnarray}
\vspace{1pc}
Further on, making use of the following equality
\vspace{1pc}
\begin{equation}
\fl
\frac{1}{\gamma^{2}_{\nu - 1,m}} \, \frac{d}{d \omega_{\alpha}} \, \left(\gamma^{2}_{\nu - 1,n} + \frac{(1- \omega_{\alpha})}{\omega_{\alpha}}\gamma^{2}_{\nu - 1,m}\right)^{-1} = \frac{1}{\left(\omega_{\alpha}  \gamma^{2}_{\nu - 1,n} + (1- \omega_{\alpha}) \gamma^{2}_{\nu - 1,m}\right)^2}
\end{equation}
\vspace{1pc} and of the  definition of the moment-generating function,
Eq. (\ref{alpha<2}), \vspace{1pc}
\begin{eqnarray}
\Phi(\sigma) &=& \int^{\infty}_0 du_{\alpha} \, P(u_{\alpha}) \, \exp\left( - \sigma \, u_{\alpha}\right) \nonumber\\
&=& \left[\Gamma\left(\nu\right) \left(\frac{\sigma}{2 - \alpha}\right)^{\frac{1-\nu}{2}} {\rm I}_{\nu - 1}\left(2 \sqrt{\frac{\sigma}{2 - \alpha}}\right)\right]^{-1} \nonumber\\
&=& \frac{2^{2-\nu}}{(2 -\alpha) \, \Gamma(\nu + 1)} \,  \sum_{n=1}^{\infty} \frac{\gamma^{\nu}_{\nu - 1,n}}{J_{\nu}\left(\gamma_{\nu-1,n}\right)} \, \left(\frac{4 \sigma}{2 - \alpha} +  \gamma^{2}_{\nu - 1,n}\right)^{-1} \,,
\end{eqnarray}
we can perform  one of the summations in  Eq. (\ref{ss}) and, finally,
expressing the Bessel functions  in terms of hypergeometric series, we
find \vspace{1pc} \begin{equation}
\label{pw2d}
\fl
P(\omega_{\alpha}) = 4 \nu \frac{d}{d\omega_{\alpha}} 
 \sum_{m=1}^{\infty} \left[ \gamma^{2}_{\nu-1,m} \,_0F_1\left(\nu + 1,-\frac{\gamma^2_{\nu-1,m}}{4}\right) \, _0F_1\left(\nu, \frac{1 - \omega_{\alpha}}{\omega_\alpha} \, \frac{\gamma^2_{\nu-1,m}}{4}\right) \right]^{-1}.
\end{equation}\vspace{1pc}  
In a similar fashion, for the case $\alpha > 2$ we obtain
\vspace{1pc}  \begin{equation}
\label{2pw2d}
\fl
P(\omega_{\alpha}) = \frac{4 (\alpha-1)}{\alpha - 2} \, \frac{d}{d\omega_{\alpha}} 
 \sum_{m=1}^{\infty} \left[ \gamma^{2}_{-\nu,m} \,_0F_1\left(2-\nu,-\frac{\gamma^2_{-\nu,m}}{4}\right) \, _0F_1\left(1-\nu, \frac{1 - \omega_{\alpha}}{\omega_\alpha} \, \frac{\gamma^2_{-\nu,m}}{4}\right) \right]^{-1}.
\end{equation}\vspace{1pc}  
One  may  readily  notice  that  these two  latter  distributions  are
normalised.

\begin{figure}[ht]
  \centerline{\includegraphics*[width=0.75\textwidth]{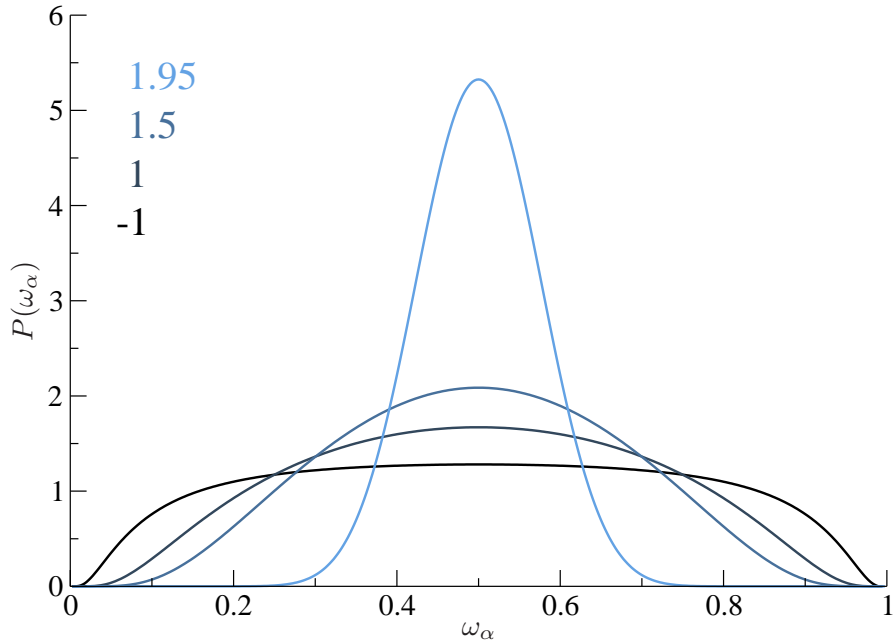}}
  \caption{(color online) The distribution $P(\omega_{\alpha})$ in
    Eq. (\ref{pw2d}) for different $\alpha < 2$ in 2D systems.}
\label{fig4}
\end{figure}

In Fig.  \ref{fig4} we plot the distribution in Eq.  \ref{pw2d} for
different values of $\alpha < 2$. One notices that similarly to the 3D
case, as $\alpha \to 2$ the distribution becomes progressively
narrower and the peak at $\omega = 1/2$ becomes more pronounced, which
means that it becomes progressively more likely that the diffusion
coefficients, deduced from two different realisations of Brownian
trajectories using the weighted least-squares estimators, will have
the same value.  One notices as well that for the unweighted LSE
($\alpha=1$), the distribution $P(\omega)$ is rather flat, pointing at
large discrepancies between the estimates obtained from different
trajectories.

\section{Conclusions}
\label{conc}

To conclude, we have studied the distribution function $P(u_{\alpha})$
of the estimators $u_{\alpha} \sim T^{-1} \int^T_0 \, \omega(t) \,
{\bf B}^2_{t} \, dt$, which optimise the least-squares fitting of the
diffusion coefficient $D_f$ of a single $d$-dimensional Brownian
trajectory ${\bf B}_{t}$.  We pursued here the optimisation further by
considering a family of weight functions of the form $\omega(t) = (t_0
+ t)^{-\alpha}$, where $t_0$ is a time lag and $\alpha$ is an
arbitrary real number, and seeking such values of $\alpha$ for which
the estimators most efficiently filter out the fluctuations.  We
calculated $P(u_{\alpha})$ exactly for arbitrary $\alpha$ and for
arbitrary spatial dimension $d$, and showed that only for $\alpha = 2$
the distribution $P(u_{\alpha})$ converges, as $\epsilon = t_0/T \to
0$, to the Dirac delta-function centered at the ensemble average value
of the estimator.  This allowed us to conclude that only the
estimators with $\alpha = 2$ possess an ergodic property, so that the
ensemble averaged diffusion coefficient can be obtained with any
necessary precision from a single trajectory data, but at the expense
of a progressively higher experimental resolution.  For any $\alpha
\neq 2$ the distribution attains, as $\epsilon \to 0$, a certain
limiting form with a finite variance, which signifies that such
estimators are not ergodic.

\end{document}